\documentclass[usenatbib]{mn2e}
\usepackage{graphicx}
\usepackage{amsmath}
\usepackage{amssymb}
\renewcommand{\text}[1]{\quad\mbox{#1}\quad}
\newcommand{\spr}[2]{\bmath{#1} \!\cdot\! \bmath{#2}}

\newcommand{\aj}{AJ}
\newcommand{\apj}{ApJ}

\newcommand{\apjl}{ApJL}
\newcommand{\mnras}{MNRAS}

\newcommand{\aap}{A\&A}

\newcommand{\apss}{Ap\&SS}
\usepackage{euscript}
\title{Accretion of a massive magnetized torus on a rotating black hole
}
\author[M.~V. Barkov \& A.~N. Baushev ]{
Maxim V.~Barkov$^{1,2}$,
Anton N. Baushev$^{3}$\thanks{E-Mail: bmv@maths.leeds.ac.uk (MVB); baushev@gmail.com (ANB)}\\
$^{1}$Department of Applied Mathematics, The University of Leeds,
Leeds, LS2 9GT\\
$^{2}$Space Research Institute, 84/32 Profsoyuznaya Street, Moscow
117997, Russia\\
$^{3}$Bogoliubov Laboratory of Theoretical Physics, Joint Institute for Nuclear Research\\
141980 Dubna, Moscow Region, Russia}
\begin{document}
\date{Received/Accepted}
\maketitle

\begin{abstract}

We present numerical simulations of the axisymmetric accretion of a massive magnetized plasma torus
on a rotating black hole. We use a realistic equation of state, which takes into account neutrino
cooling and energy loss due to nucleus dissociations. We simulated various magnetic field
configurations and torus models, both optically thick and thin for neutrinos. It is shown that the
neutrino cooling does not significantly change either the structure of the accretion flow or the
total energy release of the system. The calculations evidence heating of the wind surrounding the
collapsar by the shock waves generated at the jet-wind border. This mechanism can
give rise to a hot corona around the binary system like SS433.

Angular momentum of the accreting matter defines the time scale of the accretion. Due to the
absence of the magnetic dynamo in our calculations, the initial strength and topology of the magnetic
field determines magnetization of the black hole, jet formation properties and the total energy
yield. We estimated the total energy transformed to jets as $1.3\times 10^{52}$~{ergs} which
was sufficient to explain hypernova explosions like GRB 980425 or GRB 030329.

\end{abstract}

\begin{keywords}
gamma-rays: bursts, methods: numerical, (magnetohydrodynamics) MHD, black hole physics.
\end{keywords}

\section{Introduction}
\label{introduction}

In spite of significant progress in recent years, the nature of gamma-ray bursts (GRB), discovered
by \citet{kleb} more than 30 years ago, is still enigmatic. Although the light curves and emission
spectra of GRBs are very diverse, they seem to split into two groups of possibly different origin:
long bursts ($\Delta t > 2$~s) with a softer spectrum and short bursts ($\Delta t < 2$~s) with a
harder spectrum \citep{maz81,kouv93, fish95}.  The long GRBs are often believed to be associated
with star-formation regions \citep{bloom02,fruch06,kos05}. Only these regions can host massive
stars that have astronomically very short lifetime and die soon after the birth. In fact, recent
observations have provided strong arguments in favour of the connection of GRBs with the deaths of
massive stars.  Light curves of many GRB optical afterglows show features inherent in the
supernovae events; moreover, several long GRBs have been firmly associated with particular
supernovae, the most popular examples being GRB 980425 and SN 1998bw \citep{sof98,gal98,pian00}.
Even more convincing evidence exists in the case of the low red shift GRB 030329 ($z=0.1685$;
\citet{gre03}) and its associated supernova, SN 2003dh \citep{math03,hjo03,sok03}. The spectra of
these supernovae show exceptionally broad emission lines indicating  abnormally high velocity of
the ejecta, typical of the recently proposed ``hypernovae'' class objects.

The most popular model of the central engine of these sources is based on the ``failed supernova''
stellar collapse scenario implying that the iron core of the progenitor star forms a black hole
\citep{W93}. If the progenitor does not rotate, its collapse is likely to happen 'silently' until
all the star has been swallowed up by the black hole.  If, however, the specific angular momentum
of the equatorial part of the stellar envelope exceeds that of the last stable orbit of the black
hole, then the collapse becomes highly anisotropic.  While in the polar region it may proceed more
or less uninhibited, the equatorial layers form a dense and massive accretion disk. Then the
collapse of the layers is delayed, and the gravitational energy released in the disk can be very
large and drive GRB outflows, predominantly in the polar directions where mass density of the
accreting matter can be much lower \citep{MW99}. However, the actual process responsible for the
GRB outflows is not established and remains a subject of ongoing investigations. The main
mechanisms proposed to explain GRB outflows are neutrino pair annihilation heating
\citep{pwf99,MW99,AIMGM00}, magnetized disk wind \citep{BP82,UM06}, and magnetic braking of the
central black hole rotation \citep{BZ77,BK07}.

High-precision self-consistent models of disk dynamics and neutrino propagation are required in
order to obtain reliable results in the neutrino-driven supernova explosion theory. By now only
relatively crude calculations have been carried out, and they show that the neutrino heating need
not have a dominant role. \citet{bir07} studied the heating rate due to annihilation of neutrinos
emitted by neutrinospheres of various prescribed geometries and temperatures. The energy deposition
rates obtained in this paper lie in the range $(0.07-27)\times10^{49}/$~{erg/s}, and the typical
annihilation efficiency seems to be rather low, about $10^{-3}$. Neutrino heating from a
geometrically thin standard accretion disk \citep{ss73} is calculated in a recent article by
\citet{zb08}. It is shown that the process efficiency strongly depends on the rotation parameters
of the black hole and rapidly decreases with the distance as $r^{-4.7}$. Other aspects of the
collapsar model were considered in \citet{bau}.  \citet{shibata} carried out general relativistic
MHD simulations of accretion disks with the masses $(0.1-0.4) M_\odot$. They found that the disk
opacity for neutrinos was high, resulting in low ($\simeq 0.01 - 0.02$) efficiency of neutrino
emission itself, as most neutrinos generated in the disk could not escape from it and accreted in
the black hole. \citet{nag07} considered both neutrino heating and cooling in their Newtonian
simulations of collapsars. They concluded that the neutrino energy deposition was insufficient to
drive GRB explosions and that the magnetic mechanism was more promising.

In the last few years the role of magnetic field in driving the black hole accretion and
relativistic outflows has been a subject of active investigations via numerical simulations, that
produced numerous interesting and important results \citep{PMAB03, vhk03, krol06, mac04, mac05,
mac06, mac07, krol08a, krol08, n09, mac09}. In these studies the initial distribution described a
keplerian disk or equilibrium torus threaded with a relatively weak poloidal field, whose lines
followed the iso-density or iso-pressure contours of the disk. The disk accretion was found to be
driven by magnetic stresses via development of magneto-rotational instability (MRI)
\citep{v59,c60,bh91}. In addition to the disk, the numerical solution considered two other generic
structures - the magnetized disk corona and the highly magnetized polar funnel that hosted the
relativistic outflow from the black hole. These studies applied simple adiabatic equations of state
and did not take into account effects of radiative cooling that may be small for some types of
Active Galactic Nuclei, but not for disks of collapsing stars.

Recently \citet{shibata} have carried out two-dimensional general relativistic MHD simulations in
Kerr metrics with a realistic equation of state and taking into account the neutrino cooling, but
the physical time span of their computations was rather short, only $\simeq 0.06$~s.

In this article, we model the situation (similar to \citet{krol08a}) when a massive stellar core
collapses and the outer envelop forms a massive accreting torus. In section 2, we consider physical
processes, in section 3 --- initial conditions, in section 4 --- results, section 5 is the
discussion.

\begin{figure*}
\includegraphics[width=80mm,angle=-0]{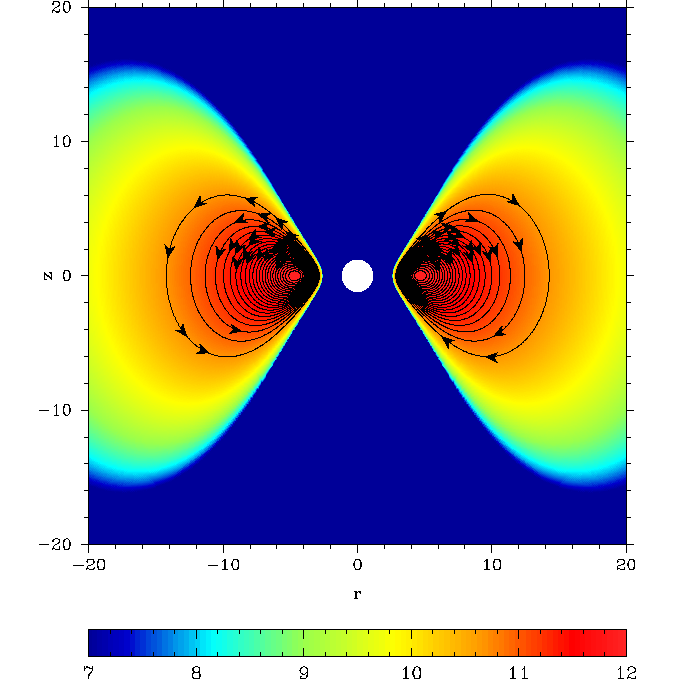}
\includegraphics[width=80mm,angle=-0]{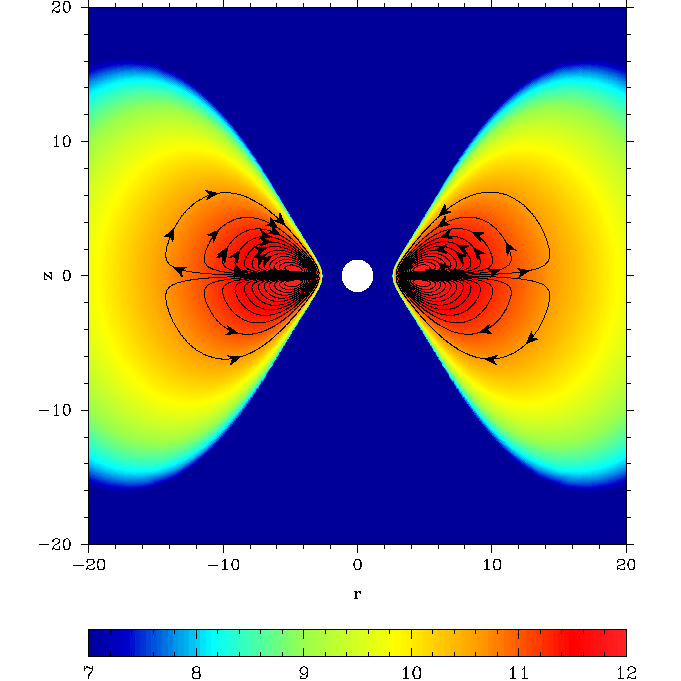}
\includegraphics[width=80mm,angle=-0]{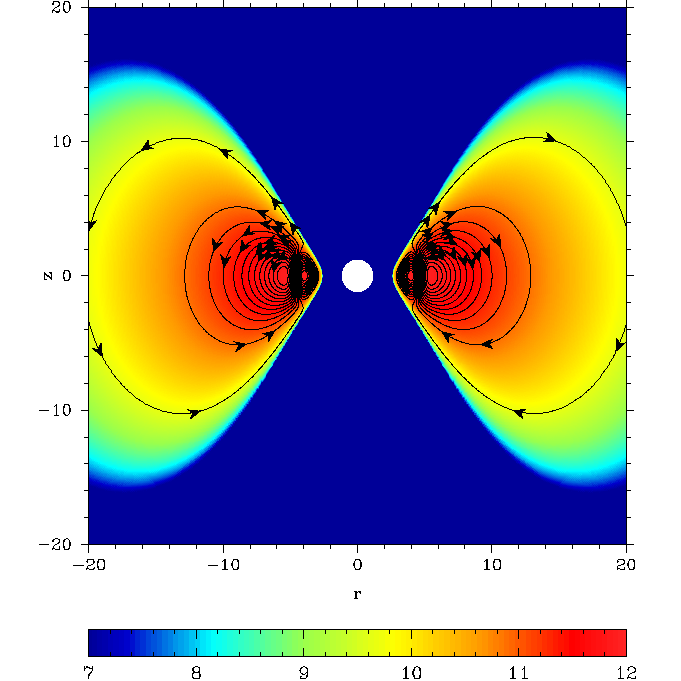}
\includegraphics[width=80mm,angle=-0]{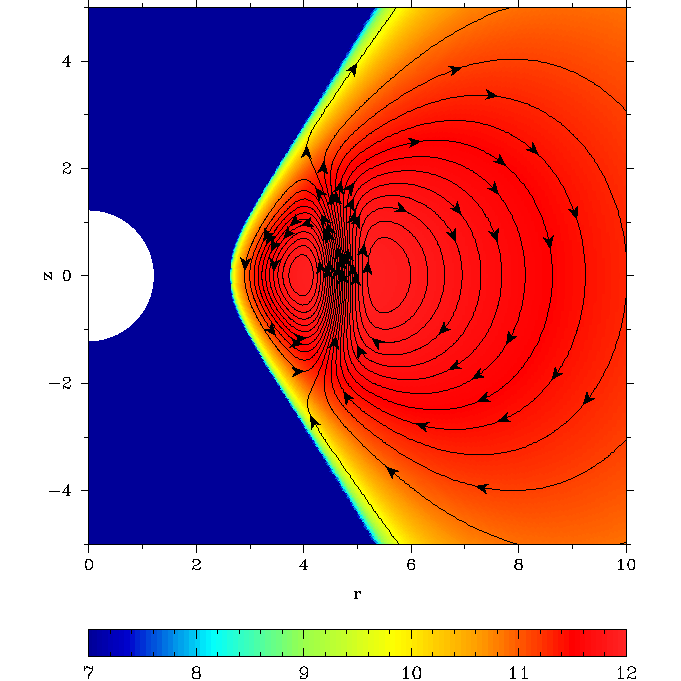}
\caption{Initial conditions for the {\it Low torus momentum} (upper left panel), {\it Quadrupole 1}
(upper right panel), and {\it Quadrupole 2} (lower panels) models. Initial distribution of the
density logarithm and the initial magnetic field are shown by colours and lines, respectively.}
\label{top_field}
\end{figure*}

\section{Physical processes}
\label{PP}

\subsection{Equation of state.}

For the simulations we use the equation of state
\begin{equation}
    P_g \equiv P(\rho,T) = P_0(\rho)+n_b k T+ \frac{\sigma T^4}{3},
\label{eos_p}
\end{equation}
where $k$ is the Boltzmann constant, $\sigma$ is the radiation energy density constant,
$n_b=\rho/m_p$ is concentration of baryons, $m_p$ is the mass of proton, $P$ is the pressure,
$\rho$ is the density and $T$ is the temperature; $P_0(\rho)$ describes the ground level for the
cold degenerate matter, the expression for $P_0(\rho)$ is an approximation of the table function
from \citet{bay71b,bay71a,mal75}.

The specific energy (per mass unit) was defined thermodynamically as
\begin{equation}
    \epsilon = \epsilon_0(\rho)+\frac{3 k T}{2 m_p}+\frac{\sigma T^4}{\rho}. \label{eos_e}
\end{equation}

The value of $\epsilon_0(\rho)$ was taken from \citet{bkbook}. This equation of state allows us to
deal with high-density matter. A more accurate EOS is significantly more expensive in calculations,
while the disregarding of the $e^+e^-$-pair component in the plasma, for instance, can give a
relative error in definition of the temperature less than $0.3$ at high temperatures
\citep{bkbook}.

\begin{figure*}
\includegraphics[width=85mm,angle=-0]{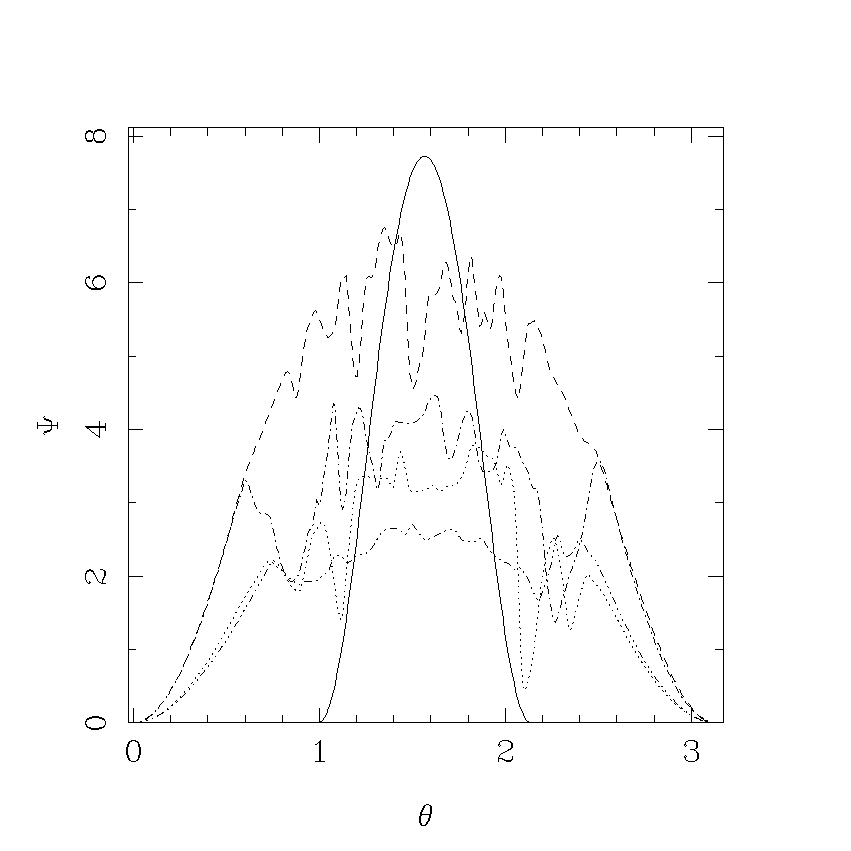}
\includegraphics[width=85mm,angle=-0]{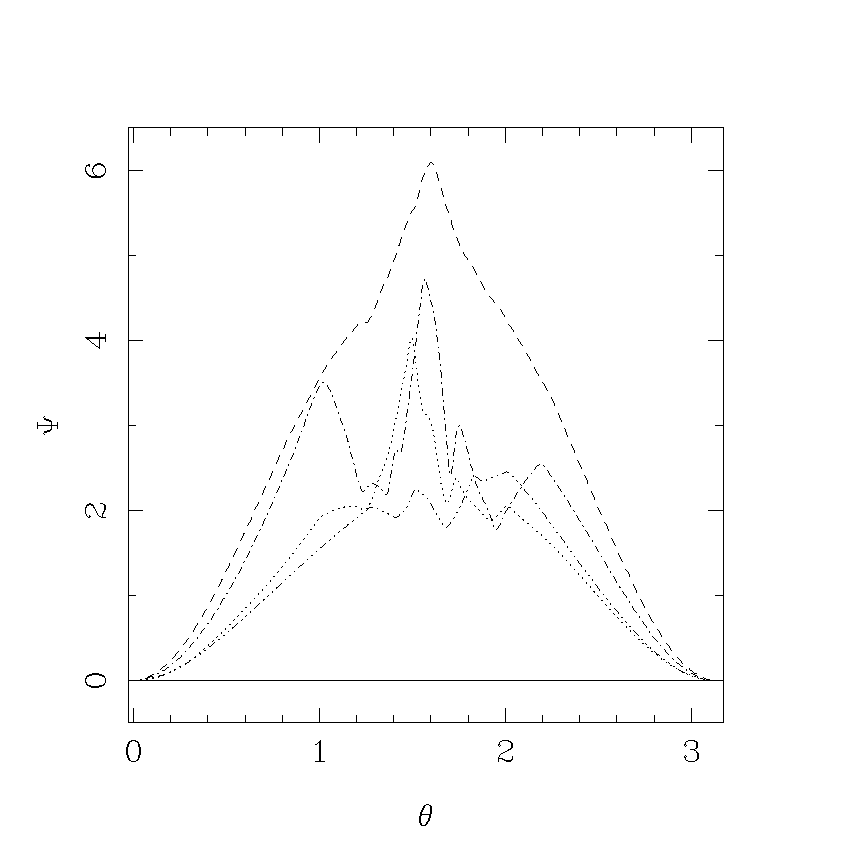}
\caption{Time evolution of the magnetic flux $\Psi$ through the sphere of radius $r=4.7 r_g$ (left
panel) and through the horizon (right panel) in the case of the {\it Low torus momentum}. The
fluxes at the moments $t=4.96\cdot 10^{-4}$~s., $t=0.0248$~s., $t=0.0495$~s., $t=0.0991$~s.,
$t=0.346$~s. are represented by solid, dashed, dot-dashed, dotted, and three-dots-dashed lines,
respectively.} \label{mag_fl}
\end{figure*}

\subsection{Neutrino cooling}

On the grounds discussed in the introduction we ignore effects of neutrino heating in our
simulations and take into account only the energy loss due to emission of neutrinos. Approximation
formulas for the cooling rates due to $e^+e^-$ pair annihilation $Q_{p}$, neutrino photo-production
$Q_{ph}$, and plasma mechanism $Q_{pl}$ are taken from \citet{schinder}, the cooling rate due to
URCA processes $Q_{u}$ --- from \citet{ivanova}, and the cooling rate due to synchrotron neutrino
emission $Q_{s}$ --- from \citet{bezchas}:
\begin{equation}
    F_{\nu}(\rho, T) = Q_{p}+Q_{ph}+Q_{pl}+Q_{u}+Q_{s}.
\label{opt}
\end{equation}
The neutrino cooling is introduced by the source term in the energy-momentum equation
 \begin{equation}
  \partial_\nu(\sqrt{-g} T^\nu_\mu)=\sqrt{-g} S u_\mu,
\label{ncool}
\end{equation}
where $u^\nu$ is the plasma four-velocity, $g$ is the determinant of the metric tensor, and $S$ is
the cooling rate as measured in the fluid frame.

One of the aims of our research was the investigation of sensitivity of the torus accretion model
to the neutrino cooling processes. We compare two cases: with no cooling at all and the optically
thin case with neutrino cooling. The first instance is the most popular \citep{gam04, mac04, mac07,
krol06, krol08} and less realistic. The second instance well suits for a light torus with the mass
of a few per cent of the solar. It could be realized when two neutron stars or a neutron star and a
black hole merge. The intermediate optical depth case is considered in \citet{B08} and can be
suitable for a collapsar model when the disk mass reaches several solar masses.

\section{Simulation setup and the description of calculational models}
\label{SS}
\subsection{General consideration}
We make our calculation in the ideal relativistic MHD approximation using an upwind conservative
scheme that is based on a linear Riemann solver and the constrained transport method to evolve the
magnetic field.  The details of this numerical method and its testing results are expounded in
\citet{K99,K04b,kom06}.

The gravitational attraction of the black hole is introduced via Kerr metric in the Kerr-Schild
coordinates, $\{\phi,r,\theta\}$. We set the black hole mass as $M_{BH} = 10 M_{\odot}$ which
corresponds to $r_g = 14.847$~{km} (we define $r_g\equiv G M_{BH}/c^2$). The two-dimensional
axisymmetric computational domain is $(r_{in}<r<r_{out}, 0\le\theta\le\pi)$, where $r_{in} \equiv
(1+\sqrt{1-a^2}/2) \; r_g$, $a\equiv c J_{BH}/G M^2_{BH}$ is the dimensionless angular momentum of
the black hole ($J_{BH}$ is the black hole angular momentum), and $r_{out} \equiv 200 r_g \simeq
2969$ km. We adopt free-outflow as the outer and the inner boundary conditions. The inner boundary
is located just inside the event horizon and adopts the free-flow boundary conditions. It is worthy
of notice that the inner boundary is inside of the outer event horizon - this choice is possible
since the horizon coordinate singularity is absent in the Kerr-Schild coordinates. The total mass
within the domain is small (less than 25\%) as compared with the mass of the black hole, which
allows us to ignore its self-gravity. The grid is uniform over $\theta$, where it has 320 cells,
and is almost uniform over $\log(r)$, where it has 459 cells, the linear size of each cell being
the same in both directions.

\begin{figure*}
\includegraphics[width=52mm,angle=-0]{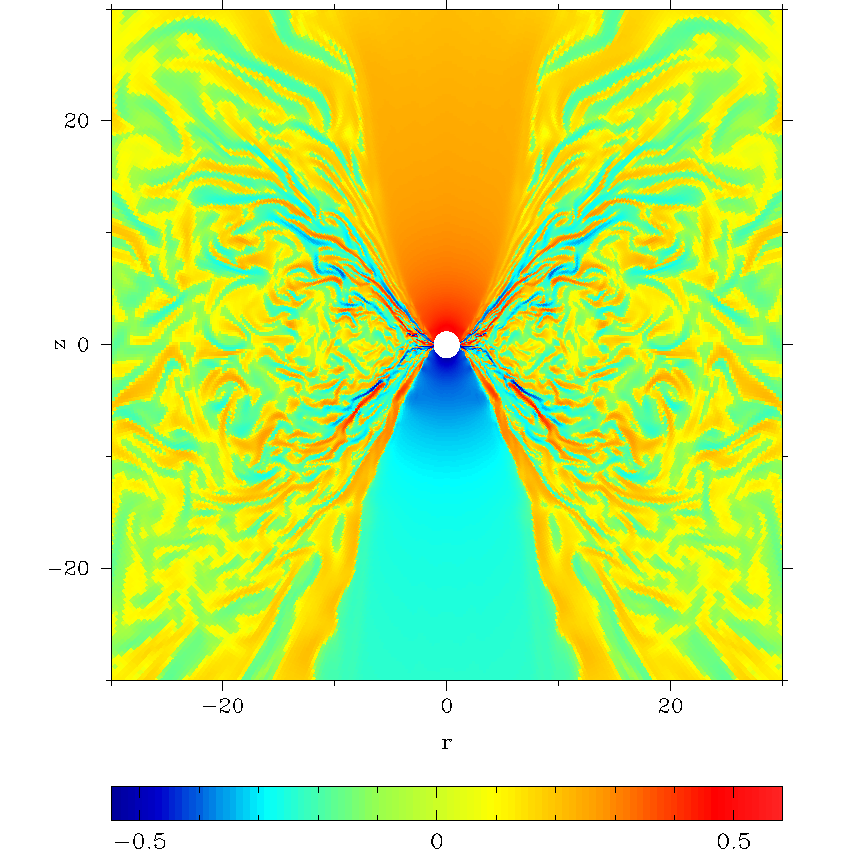}
\includegraphics[width=52mm,angle=-0]{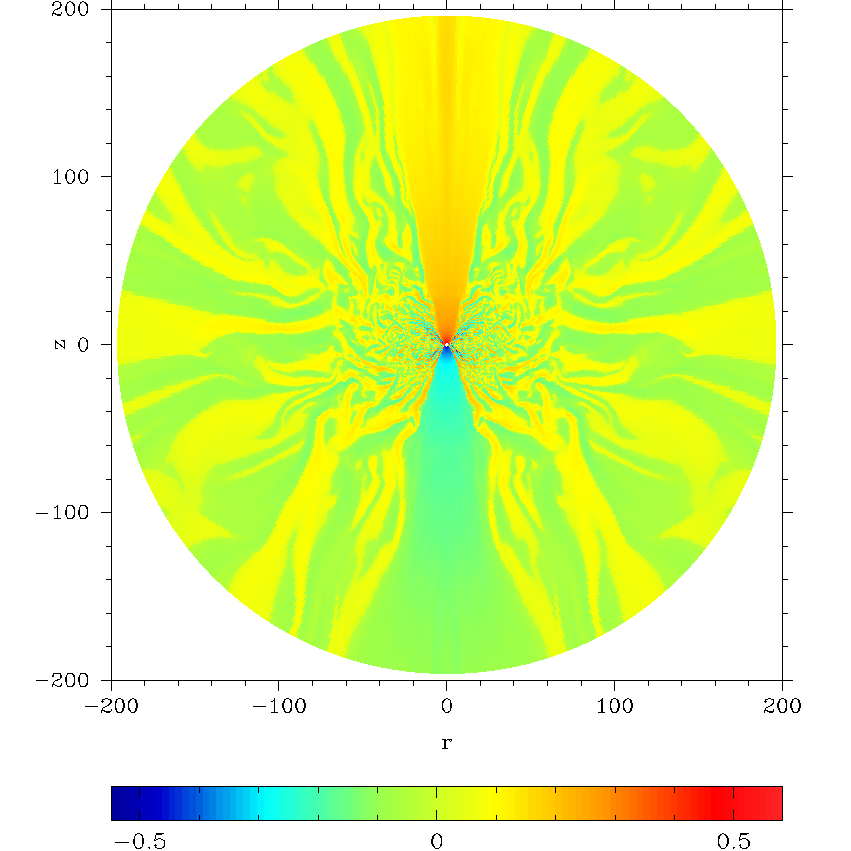}
\includegraphics[width=52mm,angle=-0]{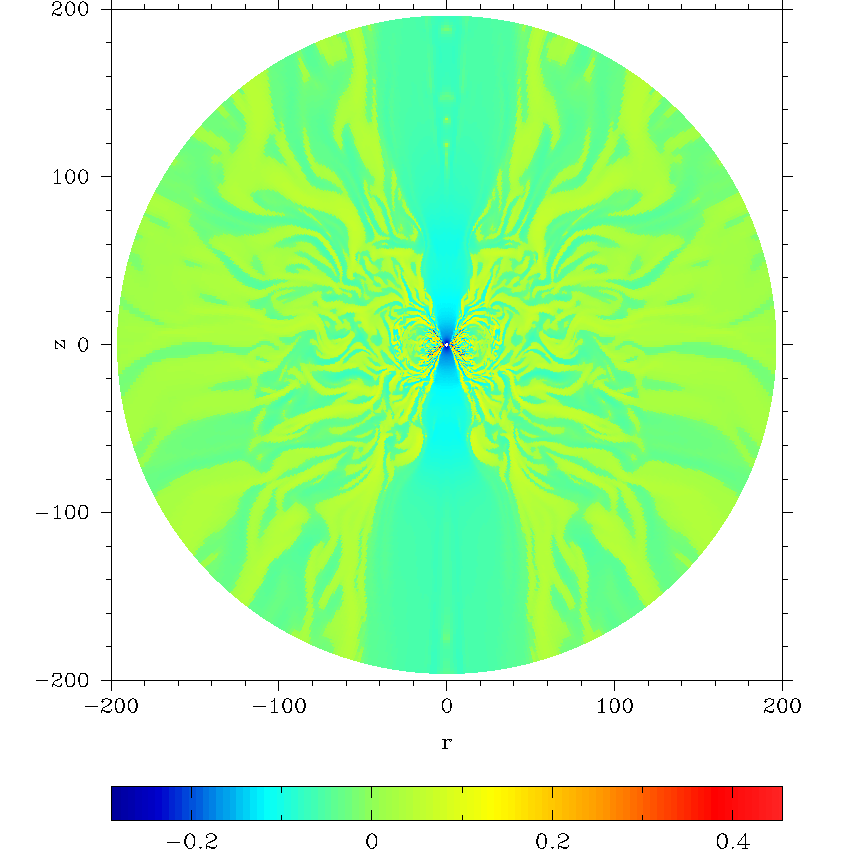}
\includegraphics[width=52mm,angle=-0]{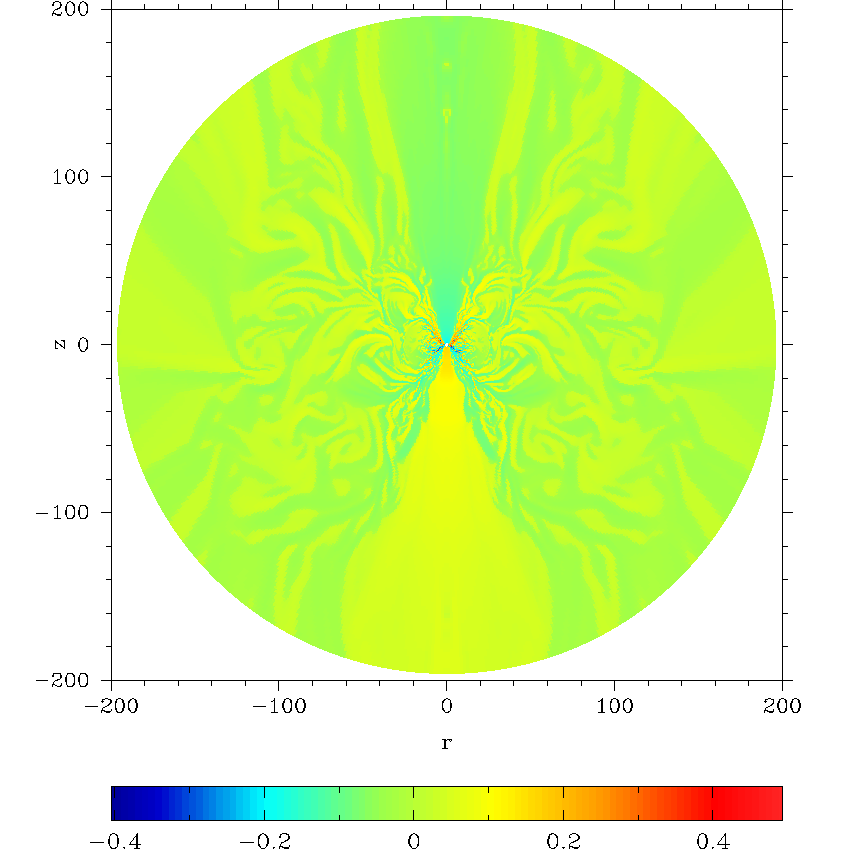}
\includegraphics[width=52mm,angle=-0]{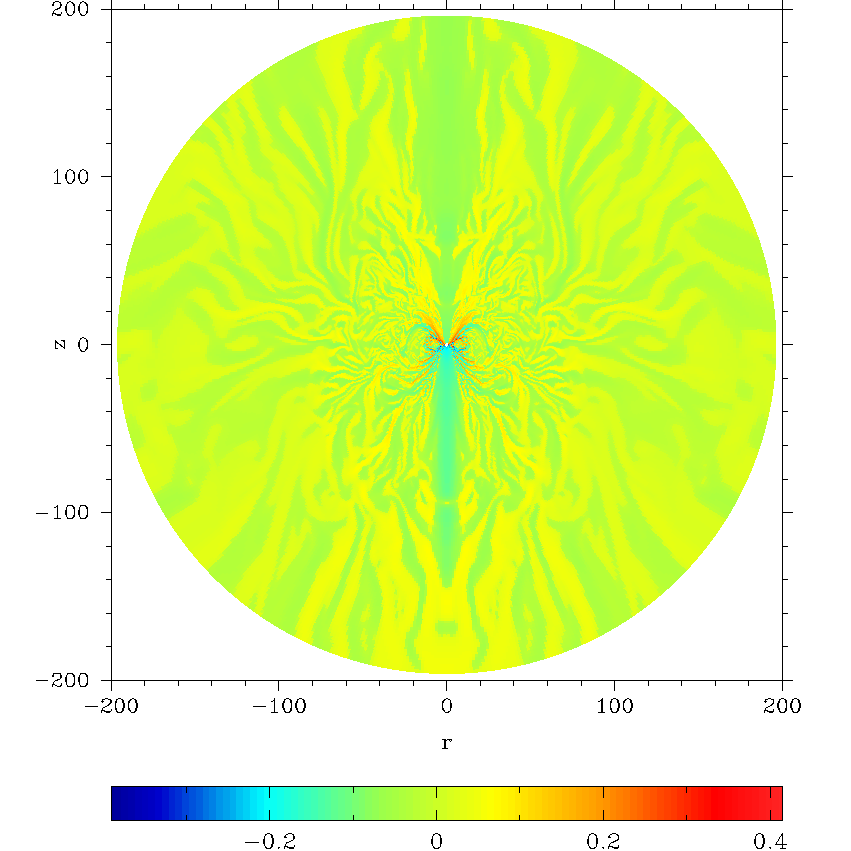}
\includegraphics[width=52mm,angle=-0]{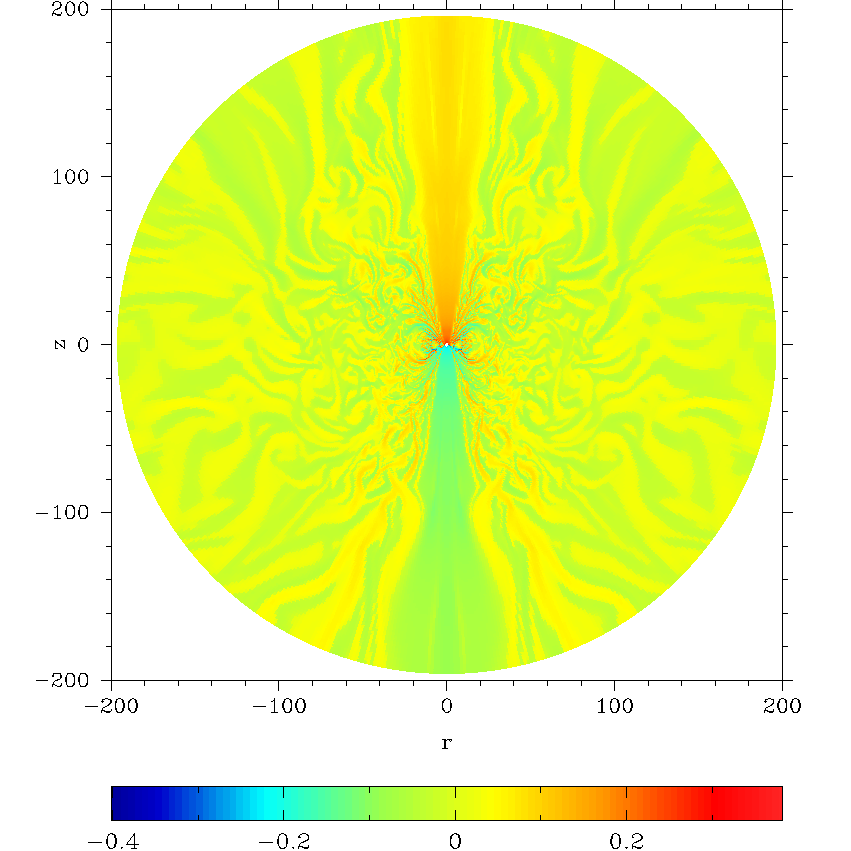}
\caption{Distribution of $sign(B_r)(|B_r|/10^{16} G)^{1/4}$ ($B_r$ is the radial component of the
magnetic field). The upper panel represents the distribution at the moment $t=0.208$~s. for the
central part of the {\it Low torus momentum} model (left picture), general view of the {\it Low
torus momentum} model (central picture), and {\it Quadrupole 1} model (right picture). The lower
panel represents the distribution for the {\it Quadrupole 2} model at the moments $t=0.1288$~s.,
$t=0.2278$~s., and $t=0.4259$~s.} \label{bf2}
\end{figure*}

As the initial distribution we consider an equilibrium torus \citep{fish76,abr78,kom06} which is a
"torus" of plasma with the black hole in the centre.  The value of the specific angular momentum
$l_0$ determines the total effective potential which can be written as

\begin{equation}
    W(r,\theta)=\frac{1}{2}\ln\left|\frac{L}{A}\right|,
\label{pot_1}
\end{equation}
where
\begin{equation}
    L=g_{t\phi}g_{t\phi}-g_{tt}g_{\phi\phi}
\end{equation}
and
\begin{equation}
    A=g_{\phi\phi}+2l_0 g_{t\phi}+l_0^2g_{tt}
\label{pot_3}
\end{equation}
Our numerical scheme cannot operate with vacuum. To avoid this difficulty, we have to introduce
minimum possible values of the density and pressure that limit how small they can be. We also check
the magnetic field and do not allow it to drop below the value defined by the equations $\rho_{min}
\leq \dfrac{1}{6} \dfrac{B^2}{8\pi c^2}$ and $p_{min} \leq \dfrac{1}{100} \dfrac{B^2}{8\pi}$. For
the magnetization typical of the task considered these inequalities are usually satisfied with a
margin of three orders. The magnetic field strength is limited by the pair production by the
neutrino-antineutrino annihilation and the Compton up scattering pair production \citep{bip92}.

\begin{figure*}
\includegraphics[width=85mm]{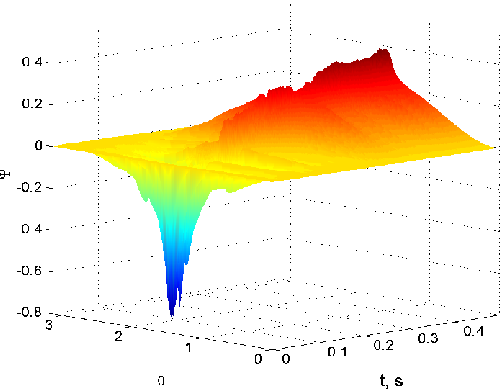}
\includegraphics[width=60mm]{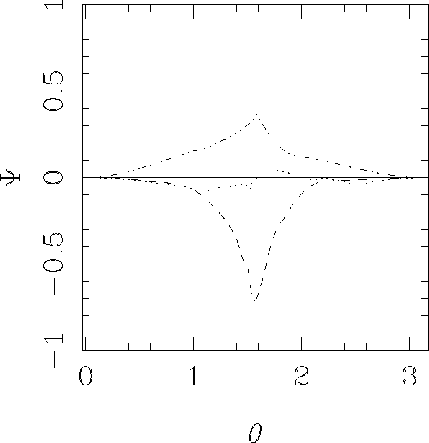}
\caption{Time evolution of the magnetic flux $\Psi$ through the horizon for the {\it Quadrupole 2}
model. The left panel represents the time sections at $t=4.96\times 10^{-4}$~s., $t=0.0248$~s.,
$t=0.1238$~s., and $t=0.4452$~s. (solid, dashed, dotted, and three-dots-dished lines,
respectively)} \label{flux_q2}
\end{figure*}

\begin{figure}
\includegraphics[width=85mm]{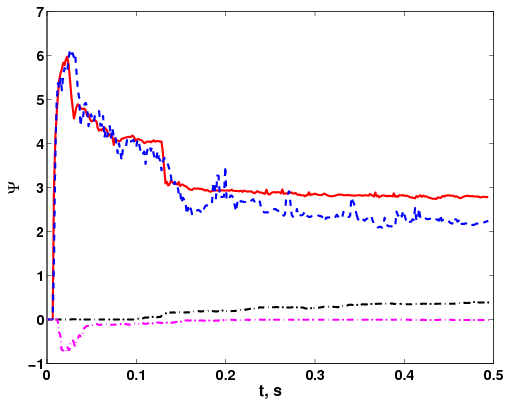}
\caption{Time evolution of the maximum of the magnetic flux $\Psi$ through the horizon for the
models: {\it Low torus momentum} (dashed line), {\it Neutrino cooling} (solid line),  and {\it
Quadrupole 2} (dot-dashed-line)} \label{flux_tq2}
\end{figure}

\subsection{Model choice}
The parameter we are most interested in is the energy release effectiveness
\begin{equation}
 \eta\equiv \frac{\int_0^T \dot E^{tot}  dt}{\int_{0}^T \dot M_{BH} c^2 dt}
\label{eta}
\end{equation}
where $\dot E^{tot}$ is the total energy flux at radius $180 r_g$ and $\dot M_{BH}$ is the
accretion rate on the horizon of the black hole. In order to test the effectiveness dependence on
the accretion conditions, we introduce several models with various initial magnetic fields and
other parameters.

The initial magnetic field structure in the disk is mainly determined by the dynamo effect. Dynamo
in accretion disks was investigated by many authors \citep{tork94,arl99,bard01} in the context of
$\alpha$-$\Omega$ model. Depending on the dynamo parameters, stationary dipole or quadrupole
magnetic fields were obtained. Some models showed time oscillations. The calculations gave only a
rough estimation of the magnetic field strength. To reproduce this simulations, we have chosen
three different topologies of the initial magnetic field (similar to \citet{krol08a}).

Due to a strong dependence of the toroidal field component on the poloidal one, we initially
introduce a purely poloidal magnetic field.  This field can be described by a vector potential with
a single nonzero component $A_{\phi}$. In our calculations we use
\begin{equation}
    A_{\phi} \propto\left\{
\begin{array}{cc}
 W(r,\theta)^3 & \; \mbox{dipole field} \\
 -W(r,\theta)^3\tanh\left(\frac{\theta-{\pi}/{2}}{w_{\theta}}\right) &
 \; \mbox{{\it Quadrupole 1} model}\\
 -W(r,\theta)^3\tanh\left(\frac{r-r_c}{w_{r}r_c}\right) & \;
 \mbox{{\it Quadrupole 2} model}
\end{array}
\right.
\label{vec_pot_3}
\end{equation}
where $w_{\theta}=0.05$ and $w_{r}=0.1$ are the parameters of the vector potential.  We adopt a
dipole initial magnetic field for {\it Low torus momentum}, {\it High torus momentum}, {\it
Neutrino cooling},  {\it Schwarzschild}, and {\it Low magnetized} models. Model {\it Quadrupole 1}
possesses a quadrupole-like field generated by two dipoles: one is above and the second is below
the equatorial plane. {\it Quadrupole 2} model also has a quadrupole field, but the two dipoles are
situated closer and farther from the black hole, which reproduces time oscillating magnetic field
in the accretion disk. The topology of magnetic field is presented in Fig.~\ref{top_field}. In all
the cases the initial field is normalized so that the maximum  of the magnetic to gas pressure
ratio $\beta=P_m/P_g$ does not exceed $3\cdot 10^{-2}$, except for the {\it Low magnetized} model,
where  $\beta \leq 3\cdot 10^{-4}$. We have chosen the small value of $\beta$ in order to avoid
strong influence of the magnetic pressure on the initial hydrodynamical equilibrium of the
configuration.

The dimensionless angular momentum of the black hole is $a=0.9$ for all the models except for the
{\it Schwarzschild} one, where $a=0$. The collapsing torus specific angular momentum and mass are
equal to $l = 2.8\cdot r_g c = 1.247\cdot 10^{17} \; cm^2 sec^{-1}, M_{tor} = 2.55 M_{\odot}$ for
{\it Low torus momentum}, {\it Neutrino cooling}, {\it Quadrupole 1}, {\it Quadrupole 2} models and
$l = 4.0\cdot r_g c = 1.782\cdot 10^{17} \; cm^2 sec^{-1}, M_{tor} = 1.87 M_{\odot}$ for {\it High
torus momentum}, {\it Schwarzschild}, {\it Low magnetized} models. In all the models, except for
{\it Neutrino cooling}, we neglect the neutrino cooling. The summary of the model parameters see in
tab.~1.

\begin{table*}
\caption{The list of model initial parameters.}
\begin{tabular}{|l|c c c c c c c c c|}
\hline model name&$l/(r_g c)$ & $a$ & $ M_{tor}$ & Neutrino cooling & Magnetic field type&
 $max(P_m/P_g)$\\
\hline
{\it Quadrupole 1} & 2.8 & 0.9 & 2.55 & No & Up-down Quadrupole & $0.03$\\
{\it Quadrupole 2} & 2.8 & 0.9 & 2.55 & No & Farther-closer Quadrupole & $0.03$ \\
{\it Low torus momentum} & 2.8 & 0.9 & 2.55 & No & Dipole & $0.03$ \\
{\it Neutrino cooling} & 2.8 & 0.9 & 2.55 & Yes  & Dipole & $0.03$\\
{\it High torus momentum} & 4.0 & 0.9 & 1.87 & No & Dipole & $0.03$\\
{\it Schwarzschild} & 4.0 & 0.0 & 1.87 & No & Dipole & $0.03$\\
{\it Low magnetized} & 4.0 & 0.9 & 1.87 & No & Dipole & $0.0003$\\
\hline

\end{tabular}

\end{table*}

\section{Results}
\label{Res}

All the models (except the {\it Schwarzschild}) after the first relaxation gave rise to the same
standard configuration: a highly magnetized jet, a thick disk and the wind from it. Let us discuss
the influence of various parameters of the system on the accretion rate and energy yield.

\subsection{Magnetic field topology}

Poloidal magnetic field plays a crucial role in the evolution and dynamics of the accretion disks.
Magneto-rotational instability (MRI) defines the accretion rate and could be the source of the
magnetic dynamo mechanism in the disk. Due to the 2D approach MRI cannot lead to the magnetic
dynamo appearance in our calculations.  It is easy to see that by the magnetic flux
\begin{equation}
 \label{baushev}
 \Psi(\theta)=\int^\theta_0 \spr{B_p}{dS}
 \end{equation} decreases with time
(Fig.~\ref{mag_fl}, {\it Low torus momentum} model). First of all, we investigate the influence on
our system of various magnetic field topologies.

A dipole-type {\it Low torus momentum} model exhibits a fast MRI growth in the disk.  At first, we
can see a powerful burst of energy extraction. It seems to be an effect of switching, after a while
the flux becomes stationary. The accretion runs actively. Magnetosphere of the black hole is
dipole-like (see Fig.~\ref{mag_fl}). The magnetic flux through the horizon is of the same order as
the initial one; the effectiveness of the total energy extraction $\eta$ is $\sim 7.4\times
10^{-4}$.

{\it Quadrupole 1} model shows a slower growth of MRI.  The magnetosphere of the black hole is
monopole-like (see Fig.~\ref{bf2}) and does not significantly change with time. The magnetic flux
through the horizon is ten times reduced, compare to the initial one that results in a low energy
extraction rate: $\eta$ is $1.1\times 10^{-5}$.

The magnetic field structure of the {\it Quadrupole 2} model actually represents a configuration of
two current frames on the different radii in the equatorial plane. In this case, the magnetosphere
is dipole-like, but time variant. One can see on Fig.~\ref{bf2} that initially the magnetic lines
near the black hole are predominantly oriented upward. As the inner currant frame is falling under
the horizon, the magnetic field lines get the opposite direction. The mean magnetic flux is also
ten times less than the initial one. The efficiency coefficient $\eta$ of the jet is $2.2\cdot
10^{-5}$.

In our calculations the accretion rate to the black hole is proportional to the maximum value of
the magnetic flux module in the torus. On the other hand, the energy release in both the quadrupole
cases drops down dramatically (up to 250 times, see tab.~2).

In the case of the {\it Low magnetized} model, the MRI grows very slowly. Accretion starts after a
long ($0.6$~{sec}) phase of the field linear amplification in the torus up to $(P_m/P_g) \sim 1$.
When the accretion starts, the system takes on the standard appearance: a jet, an accretion disk
and the wind from it. The energy release and the accretion rate are very low (see tab.~2). The
efficiency $\eta$ of the wind is $2.0\times 10^{-5}$ and of the jet is $4.9\times 10^{-7}$.

\subsection{Angular momenta of the torus and of the black hole}

In the models {\it Low torus momentum}, {\it Quadrupole 1}, and {\it Quadrupole 2} we put the torus
angular momentum  $l=2.8\cdot r_g c$ which is very close to the limit configuration, and even small
perturbations lead to accretion. If the initial momentum is augmented to $l=4.0 \cdot r_g c$ ({\it
High torus momentum} model), it decreases the accretion rate thirty times but the energy release
does not drop so strongly (see tab.~2), since a significant part of the field accretes on the black
hole increasing the magnetic flux near it. As a result, the effectiveness of energy extraction in
this case is high $\eta=3.0\times10^{-3}$.

\begin{figure*}
\includegraphics[width=84mm,angle=0]{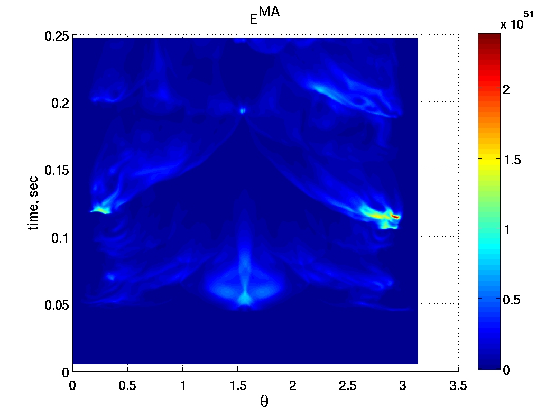}
\includegraphics[width=84mm,angle=0]{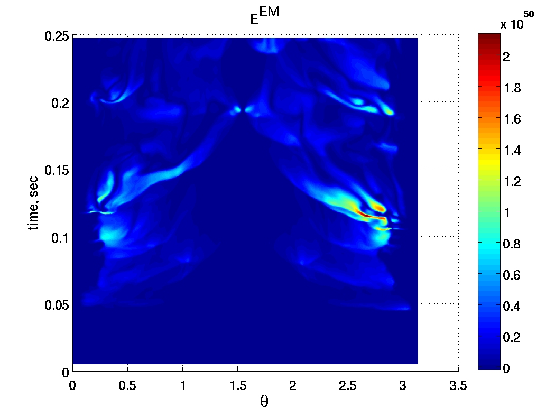}
\includegraphics[width=84mm,angle=0]{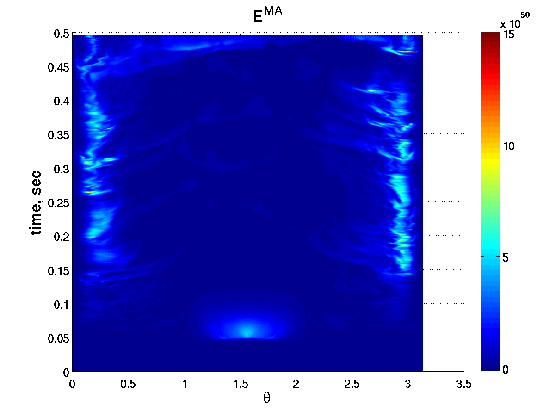}
\includegraphics[width=84mm,angle=0]{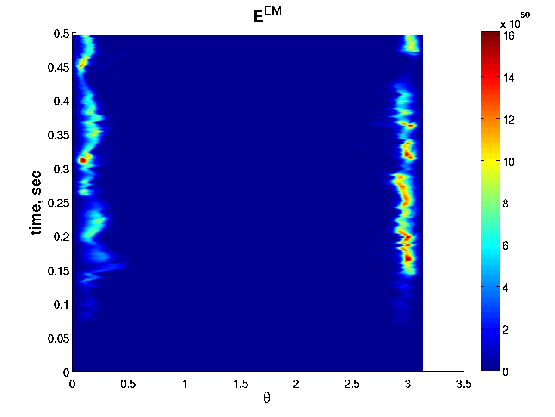}
\caption{Dependence on $\theta$ and time of the matter ($\dot E^{MA}$, left panels) and
electromagnetic ($\dot E^{EM}$, right panels) energy fluxes per unit angle at the radius $R=180
r_g$ for the {\it Schwarzschild} (upper row) and {\it High torus momentum} (lower row) models.}
\label{fl_sl}
\end{figure*}

Another important parameter is the angular momentum of the black hole $a$. Detailed investigations
of its influence on the jet formation has been made in \citet{gam04,mac04,mac05,krol06}. The
structure of the solution significantly changes if $a = 0$ ({\it Schwarzschild} model, see tab.~2).
The central highly magnetized jet does not appear at all, the energy is released through the 
subrelativistic magnetized wind. The total energy flux of the system jet-wind is almost the same
for the {\it Schwarzschild} and the {\it High torus momentum} models, whereas the accretion rate in
the former case is $27$ times higher.

Let us consider the time-space evolution of the energy flux at the distance $R=180\cdot r_g$. The
total flux can be divided in two parts: electromagnetic ${\dot{E}}^{EM}$ and carried out by matter
${\dot{E}}^{MA}$. In the case of {\it High torus momentum} (Fig.~\ref{fl_sl}, the lower row), it is
easy to see that the flux is predominantly electromagnetic in the jets and matter-dominated in the
wind. The maximal electromagnetic flux propagates near (not on) the axes. The maxima of the matter
energy flux are strongly correlated with the electromagnetic maxima, as the main part of matter
energy flux is driven by shock waves appearing at the boundary between the relativistic jets and
the non-relativistic wind (the lines going up from the jet to the equatorial region). The jets very
effectively heat the surrounding wind and can produce a hot corona around the disk.

In the case of {\it Schwarzschild} model, the situation is different (fig.\ref{fl_sl}, the upper
row): electromagnetic and matter energy fluxes have much wider angular distribution. A highly
magnetized zone does not appear, and we can see a matter dominated magnetized outflow. Magnetic
braking of the black hole produces the wind that carries out the magnetic field.

\subsection{Neutrino cooling}

In the initial torus configuration radiation and degenerated matter produce $80\%$ and $20\%$ of
the total pressure, respectively. Therefore, neutrino cooling becomes very important.

\begin{figure*}
\includegraphics[width=56mm,angle=-90]{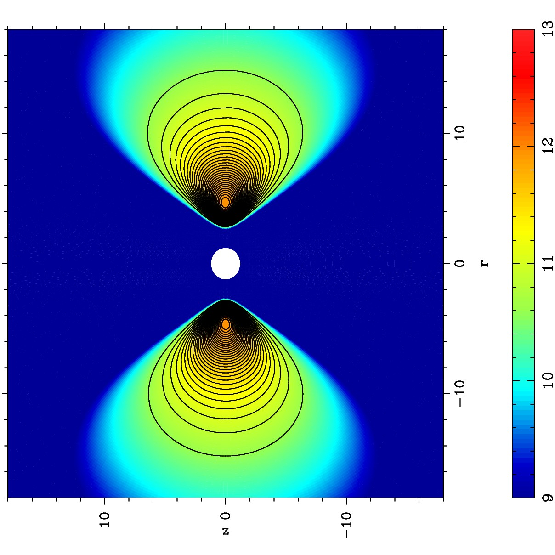}
\includegraphics[width=56mm,angle=-90]{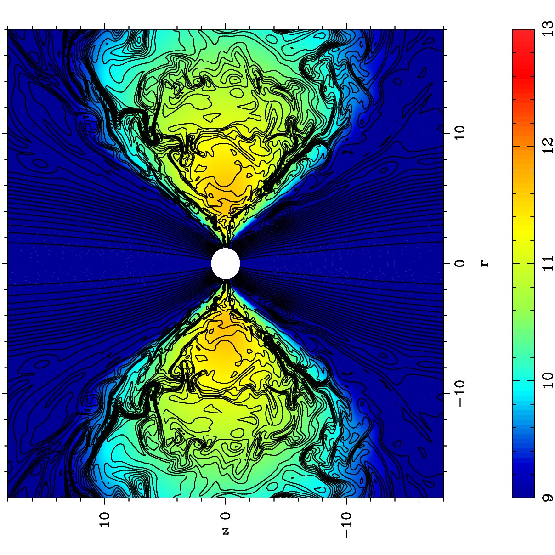}
\includegraphics[width=56mm,angle=-90]{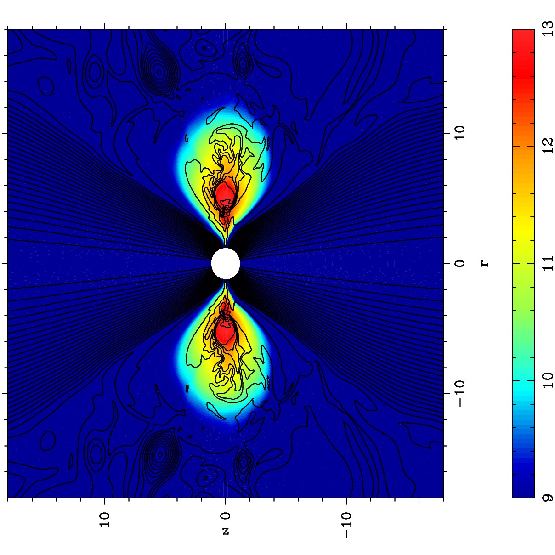}
\caption{Distribution of $\log_{10}(\rho)$ and magnetic lines. Left panel represents the initial
distribution. Central and right panels represent the distributions after 0.2075~s. of evolution for
the {\it Low torus momentum} and {\it Neutrino cooling} models are shown on the central and right
panel, respectively.} \label{rho_c}
\end{figure*}

Intensive neutrino cooling in the {\it Neutrino cooling} model (where the system is not opaque for
neutrino propagation) leads to a significant disbalance in the initial distribution. The originally
stable torus collapses to a new configuration with negligible thermal pressure. One can see the
influence of neutrino cooling on the structure of the torus in Fig.~\ref{rho_c}. The torus without
cooling keeps the initial size and shape. The cooling leads to a collapse of the initial
configuration. The maximum density increases up to ten times and reaches the value
{$10^{13}$~g/cm$^3$}.

\begin{figure*}
\includegraphics[width=84mm,angle=-0]{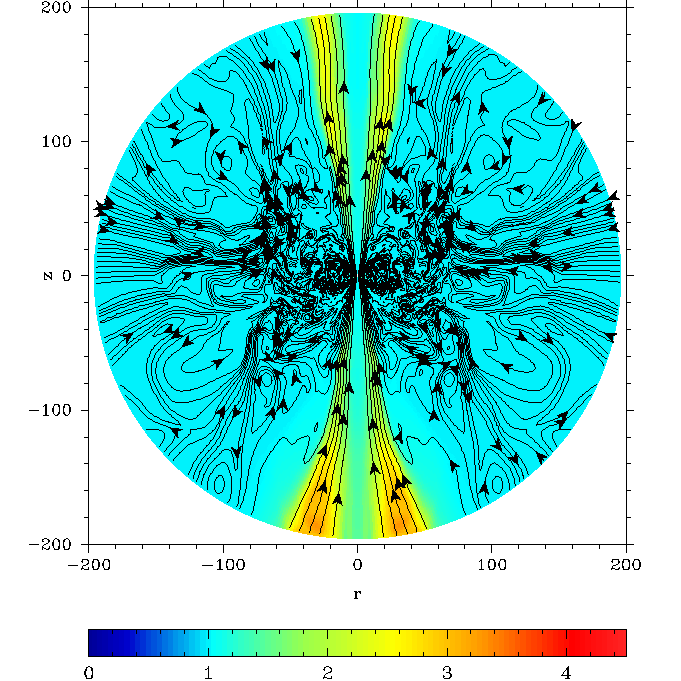}
\includegraphics[width=84mm,angle=-0]{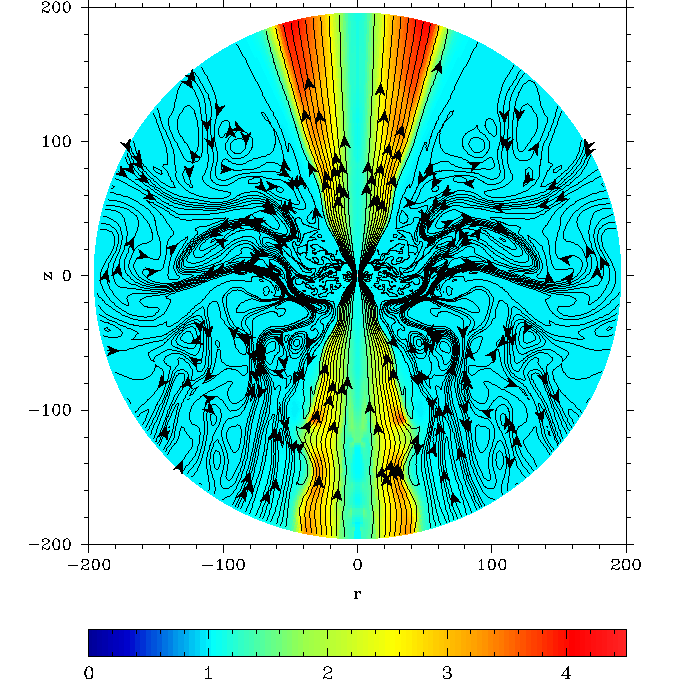}
\caption{Lorenz factor distribution for $t=0.2075$~s.  Left panel represents the {\it Low torus
momentum} model with no cooling, the right one --- {\it Neutrino cooling} model.} \label{lor_c}
\end{figure*}

The structure of the accretion flow does not change significantly: the {\it Neutrino cooling} model
flow is similar to {\it High torus momentum} (Fig.~\ref{rho_c}) and thinner than in {\it Low torus
momentum} model. The {\it Low torus momentum} model has a much smaller and less stable highly
magnetized jet region than in the {\it Neutrino cooling} case (Fig.~\ref{lor_c}). Despite the fact
that the {\it Neutrino cooling} accretion rate is two times lower than the {\it Low torus momentum}
one (tab.~2), the energy release and the magnetic flux through the horizon are even slightly higher
in the {\it Neutrino cooling} model (Fig.~\ref{flux_tq2}). So {\it Neutrino cooling} and {\it High
torus momentum} models have the same effectiveness coefficient $\eta\sim 0.003$ that is $2 - 4$
times greater than for the {\it Low torus momentum} model.

\begin{figure}
\includegraphics[width=84mm,angle=0]{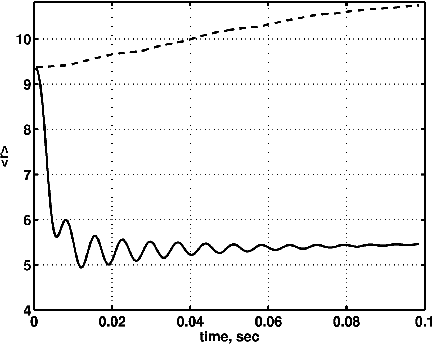}
\caption{Mean radius $r^m$ evolution for {\it Low torus momentum} (dashed line) and  {\it Neutrino
cooling} (solid line) models.} \label{oscil}
\end{figure}

At the inception of the calculations the central region of the torus rapidly cools down and falls
to the potential hollow, while the outer parts do not cool at first. Intensive neutrino cooling
leads to a disbalance in the accretion torus. Due to asymmetry of the potential well, uncompensated
gravitational forces give rise to oscillations (Fig.~\ref{oscil}, similar results were obtained by
\citet{zan05,mon07}). So the cooling produces intensive radiative shock waves, that (as well as the
poloidal magnetic field) strongly increases the effective viscosity. When we neglect the cooling,
the oscillations do not appear, the mean radius increases due to the strong wind outflow.  In the
optically thin case, the oscillations appear with the amplitude ${\Delta r^m/r^m}=0.05$, where $r^m
= \int_M |r|dm / \int_M dm$ is the mean mass radius, and quickly degrade due to the high viscosity.
We can estimate the relaxation time as $t_{relax}=t_{osc}\times Q$, where $Q=|(\Delta r_i^m+\Delta
r^m_{i+1})/2(\Delta r^m_i-\Delta r^m_{i+1})|\simeq 3.8$ and $\Delta r^m_i$ is the maximal deviation
from the equilibrium of the $i$-th oscillation.  In our case $t_{relax}\approx 0.027$~{sec}. The
same oscillations were obtained  in the case of intermediate neutrino optical depth \citep{B08}.

\begin{table*}
\caption{Main results.}
\begin{tabular}{|l|c c c c c c c c c|}
\hline model & $l$ & $a$ & $\dot M_{BH}$ & $\dot M_{w} $ &
 ${\dot{E}}^{EM}_{51}$ & ${\dot{E}}^{MA}_{51}$ & ${\dot{E}}^{tot}_{51}$ &
$\eta$ &\\
 \hline
{\it Low torus momentum} & 2.8 & 0.9 & 1.3937 & 0.0189 & 0.777 & 1.076 & 1.8530 &
 $7.4\times10^{-4}$\\
{\it Neutrino cooling} & 2.8 & 0.9 & 0.8588 & 0.0134 & 1.5024 & 1.4014 & 2.9039 &
$1.9\times10^{-3}$\\
{\it Quadrupole 1} & 2.8 & 0.9 & 0.3976 & 0.00023 & 0.0022 & 0.0058 & 0.0080 &
$1.1\times10^{-5}$\\
{\it Quadrupole 2} & 2.8 & 0.9 & 0.2126 & 0.00034 & 0.0019 & 0.0065 & 0.0083 &
$2.2\times10^{-5}$\\
{\it High torus momentum} & 4.0 & 0.9 & 0.053 & 0.0026 & 0.1283 & 0.1541 & 0.2824 &
$3.0\times10^{-3}$ \\
{\it Schwarzschild} model & 4.0 & 0.0 & 1.411 & 0.0086 & 0.0216 & 0.2320 & 0.2536 &
$1.0\times10^{-4}$ \\
{\it Low magnetized} & 4.0 & 0.9 & 0.0023& $5.5\times10^{-6}$ &
 $2.0\times10^{-6}$ & $8.16\times10^{-5}$ & $8.36\times10^{-5}$ &
 $2.0/0.049\times10^{-5}$
\end{tabular}

\medskip

Here $l$ is the dimensionless angular momentum of the accreting matter; ${\dot{M}_{BH}}$ ---
accretion rate $M_{\odot} \mbox{s}^{-1}$; $\dot M_{w} $ -- wind mass loss rate $M_{\odot} \mbox{
s}^{-1}$ at radius $180 r_g$; ${\dot{E}}^{EM}_{51}$, ${\dot{E}}^{MA}_{51}$, ${\dot{E}}^{tot}_{51}$
are electromagnetic energy, matter energy, and the total energy fluxes at radius $180 r_g$,
$10^{51} \mbox{ erg } \mbox{s}^{-1}$ ; $\eta$ --- accretion effectiveness (\ref{eta}) (for the {\it
Low magnetized} model we present separately the effectiveness of the electromagnetic energy
extraction). Notice that in the case of the {\it Low magnetized} model the accretion starts only
after $0.75$~s., while the calculation ends at $0.99$~s.

\end{table*}

\section{Discussion}
\label{Dis} Topology and strength of the magnetic field play a crucial role in the process of
accretion on a black hole.  In our simulations the energy release effectiveness varies over two
orders of magnitude when we change the magnetic field topology from the dipole ({\it Neutrino
cooling}, {\it High torus momentum} models) to the quadrupole one({\it Quadrupole 1}, {\it
Quadrupole 2} models). Due to the 2D approach we cannot consider the dynamo effect. However, we can
see temporal increases of the magnetic field intensity owing to the MRI-driven turbulence growth
and magnetic line twist.

The Blandford-Znajek mechanism plays a leading role in accreting black hole energy release. For a
force-free monopole magnetosphere the Blandford-Znajek power is given by
\begin{equation}
 \dot{E}_{BZ}=\frac{1}{6c}\left(\frac{\Omega\Psi}{4\pi}\right)^2,
\label{ebz}
\end{equation}
We assume $\Omega=\Omega_h/2$, were $\Omega_h=a/2(1+\sqrt{1-a^2})$ is the angular velocity of the
black hole and $\Psi$ is the magnetic flux (\ref{baushev}) threading the black hole.

Our estimations of the energy release effectiveness differ by a factor of $3$ from \citet{mac05}
and by a factor of up to $25$ from \citet{BK08b,KB09}. We explain it by distinctions between the
initial magnetic fields: in \citet{KB09}, for instance, the magnetic field flux was $4.8$ times
higher.

Neutrino cooling processes significantly suppress the wind from the accretion disk, but do not
diminish appreciably the energy release; moreover, sometimes they can stabilize the outflow. The
neutrino driven oscillations (Fig.~\ref{oscil}) look like an artificial effect of turning on the
initial conditions: more realistic conditions like \citet{mcf99,BK07} do not lead to significant
radial oscillations. But even if the effect is real, the oscillations \citep{mcf99} can  scarcely
be observed by gravitational wave detectors.

Highly magnetized jets appear in all the models considered, where the black hole possesses angular
momentum. The obtained parameters of the plasma acceleration in the jets (Fig.~\ref{lor_c}) are in
good agreement with \citet{KBVK07,BK08b,KVKB08}.

The boundary between the jet and the wind is unstable (the perturbation growth is most pronounced
in Fig.~\ref{lor_c}, {\it Neutrino cooling} case, in the south hemisphere in the radius interval
$100-150 r_g$. \citet{mac06} reported about a similar effect). The jet speed is much higher than
the sound speed in the wind, and the boundary perturbations generate strong shock waves propagating
to the wind region (Fig.\ref{fl_sl}).  The shocks effectively heat the wind and can produce a hot
corona around the accretion disks like in the SS433 system \citep{cher06}.

If we adopt the effectiveness coefficient $\eta\sim 0.003$ of the models {\it Neutrino cooling} and
{\it High torus momentum},  it allows us to estimate the total energy release of the torus
accretion as $E_{tot} \approx \eta M_{torus}c^2 \approx 1.3 \cdot 10^{52}$~{erg}. This value of
$\eta$ however, need not to be universal: as we saw, other magnetic field topologies can have a
significantly different efficiency.

The scenario considered in this article is a possible alternative to the magnetar-driven hypernova
scheme \citep{KB07}. Both the scenarios predict the appearance of collimated jets energetic enough
to explain GRB as well as hypernovae phenomena. Accurate calculations are needed to find
characteristic features and distinguish these two schemes. By now we can say that the chemical
composition of matter in the jets should be different. In the collapsar scenario we expect a highly
Poynting-dominated jet ($\epsilon_{field}/\epsilon_{matt}\sim 1000$, where $\epsilon_{field}$ and
$\epsilon_{matt}$ are the field and the matter energy flux densities, respectively), mainly made up
of $e^+e^-$-pairs, while in the magnetar-driven scenario the jet is baryon-dominated and
intermediately magnetized ($\epsilon_{field}/\epsilon_{matt}\sim 1-100$).

The effect of the magnetic field sign change found in the jet of the {\it
Quadrupole 2} model, may be the key to understanding of the "surprising evolution of the parsec-scale
Faraday rotation gradients in the jet of the BL Lac object B1802+784" \citep{mahgb09}. Complicated topology
of the magnetic field in the accretion disk can lead to variations of the magnetic field direction in the jet.
In this case, the jet constitutes a chain of regions with various magnetic field orientations.

\section{Conclusion}

The accretion rate and energy release are governed by four main parameters: magnetic field
structure, angular momentum of the black hole, specific angular momentum of the accreting matter,
accretion disk mass or accreting part of the star envelop. To have a long-time accretion, angular
momentum of the accreting matter should be high enough to form an accretion disk around the black
hole. As it has been demonstrated by the {\it Schwarzschild} model, if the black hole has deficient
angular momentum, the powerful magnetized jet does not appear at all, which totally changes the
picture of the accretion. Moreover, a comparison of the {\it Low} and {\it High torus momentum}
models shows us that a respectively small variation of the angular momentum of the accreting matter
leads to a change of the accretion efficiency. Neutrino losses do not prevent energy release and
can even stabilize the magnetized jet outflow.

The least understood and a very important factor defining the accretion efficiency is the disk
magnetic field structure. As we could see from the calculations, a mere replacing of the dipole
field by the quadrupole one decreases the energy release by two orders of magnitude. The magnetic
flux $\Psi$ governs the energy extraction from ergosphere (\ref{ebz}), while the black hole
magnetosphere is formed and supported by the disk. In our calculations we preset the initial
magnetic field in the torus. It is clear, however, that self-consistent numerical calculations of
the magnetic dynamo in the disk are necessary in order to elucidate the problem.

Nevertheless, our calculations show that a very intensive electromagnetic and matter energy flux
can be evolved during massive torus accretion on a rotating black hole with mass $\sim 10
M_{\odot}$. The total energy budget is quite sufficient to explain hypernova explosions like GRB
980425 or GRB 030329. The magnetized matter accretion on a rotating black hole is a very promising
model of the central engine of these phenomena.

In addition, we have found that the instability of the boundary between the jet and the wind
generates shock waves propagating to the wind region and heating it. This effect can give rise to a
hot corona around the binary system  similar to SS433.

\section*{Acknowledgments}
We would like to express our deep gratitude to Prof. S.~S.~Komissarov for helping in the
computational code creation, and for very fruitful discussions. We thank Prof.
G.~S.~Bisnovatyi-Kogan for helpful discussions.

The calculations were fulfilled at the St. Andrews UK MHD cluster and at the White Rose Grid
facilities (Everest cluster).  This work was supported by the PPARC (grant ``Theoretical
Astrophysics in Leeds'') and by the RFBR (Russian Foundation for Basic Research, Grant
08-02-00856).


\end{document}